\documentclass[conference]{IEEEtran}
\IEEEoverridecommandlockouts
\usepackage{cite}
\usepackage{amsmath,amssymb,amsfonts}
\usepackage{mathrsfs}
\usepackage{algorithmic}
\usepackage{graphicx}
\usepackage{subfigure}
\usepackage{textcomp}
\usepackage{amssymb}
\usepackage{xcolor}
\usepackage{threeparttable}
\usepackage{multirow}
\usepackage{balance}
\usepackage{pifont}
\def\BibTeX{{\rm B\kern-.05em{\sc i\kern-.025em b}\kern-.08em
    T\kern-.1667em\lower.7ex\hbox{E}\kern-.125emX}}
\begin{document}

\title{CDEdit: A Highly Applicable Redactable Blockchain with Controllable Editing Privilege and Diversified Editing Types\\
}

\author{\IEEEauthorblockN{ Xiaofeng Chen}
\IEEEauthorblockA{\textit{School of Cyber Science and Technology} \\
\textit{Beihang University, China}\\
Email: cryptocxf@buaa.edu.cn}
\and
\IEEEauthorblockN{Ying Gao\textsuperscript{*}}
\IEEEauthorblockA{\textit{School of Cyber Science and Technology} \\
\textit{Beihang University, China}\\
\textit{Key Laboratory of Aerospace Network Security, }\\
\textit{Ministry of Industry and Information Technology, China}\\
Email: gaoying@buaa.edu.cn}
}

\maketitle

\begin{abstract}
Redactable blockchains allow modifiers or voting committees with modification privileges to edit the data on the chain. Trapdoor holders in chameleon-based hash redactable blockchains can quickly compute hash collisions for arbitrary data, and without breaking the link of the hash-chain. However, chameleon-based hash redactable blockchain schemes have difficulty solving the problem of multi-level editing requests and competing for modification privileges. In this paper, we propose CDEdit, a highly applicable redactable blockchain with controllable editing privilege and diversified editing types. The proposed scheme increases the cost of invalid or malicious requests by paying the deposit on each edit request. At the same time, the editing privilege is subdivided into request, modification, and verification privileges, and the modification privilege token is distributed efficiently to prevent the abuse of the modification privilege and collusion attacks. We use chameleon hashes with ephemeral trapdoor (CHET) and ciphertext policy attribute-based encryption (CP-ABE) to implement two editing types of transaction-level and block-level, and present a practical instantiation and security analysis. Finally, the implementation and evaluation show that our scheme only costs low-performance overhead and is suitable for multi-level editing requests and modification privilege competition scenarios.
\end{abstract}


\section{Introduction}
A blockchain is a hash-chain formed by the hash of the block linking the next block in chronological order. This hash-chain linked by hash values makes the transaction content immutable and improves the trustworthiness of the transaction information. On the other hand, the immutability of transactions has become a new obstacle limiting blockchain applications.
	
In an immutable blockchain, any transaction written by a participant cannot be modified. Some malicious participants may post information that contains sensitive or illegal data. For instance, materials infringe on intellectual rights [1] and child sexual abuse images [2]. This sensitive information cannot be modified after being uploaded, which will have a severe and irreversible impact on their lives. Moreover, the General Data Protection Regulation (GDPR) [3] issued by the European Union in 2018 states that any data subject must request the responsible party to hide or delete data records about individuals. Thus, there is a significant need to edit the data on the blockchain in practice.
	
Nowadays, redactable blockchains are mainly implemented by the chameleon hash (CH) algorithm [4] [5], which uses CH with a trapdoor key to replace the traditional collision-resistant hash algorithm. In the editing mechanism with a CH algorithm, the trapdoor holder can find a collision in a broad sense without changing the hash output and breaking the hash link. According to the different data editing targets, there are two redactable blockchains. Type one is coarse-grained block-level (Bl-level) editing, which enables modification of the entire block by replacing the traditional Merkle root hash with a CH. Type two is a fine-grained transaction-level (Tx-level) edit operation designed to modify a specific transaction, more aligned with realistic requirements. The PCHBA scheme [6] links the modified transactions to the responsible modifier and does not impose a privilege penalty on the modifier who abuses rewriting privilege. It does not prevent the irresponsible modifier from abusing the privilege again the next time they edit. Although the KERB scheme [7] uses the deposit deduction method to hold malicious modders accountable and punishable. However, it does not reduce the modification privileges of malicious modifiers, resulting in the inability to eliminate malicious behaviour. To the best of our knowledge, most redactable blockchain protocols suffer from two problems.
\begin{itemize}
	\item It is challenging to manage and distribute the appropriate editing privileges. The use of centralized editing privilege control is subject to the risk of centralization [8], while the management of editing privileges in multiple centres leads to being very inefficient.
	\item It is unable to handle edit requests with different granularity or conflicts. Inefficient or ambiguous one-by-one processing consumes system resources. It leads to security issues such as conflicting data on-chain.
\end{itemize}
	
\textbf{This work.} This paper is inspired by [6] and [7]. We propose a highly applicable redactable blockchain with controllable editing privilege and diversified editing types (CDEdit). CDEdit enables multi-level editing of data on-chain in permission settings (e.g. Hyperledger [9]). The main contributions are summarized as follows.
  
The main contributions are summarized as follows.
\begin{itemize}
	\item [1.]
	\textbf{Token-based controlled editing privileges.} We propose a mechanism to manage and distribute token-based editing privileges, in which privileged tokens at different granularity are generated by a privileged token service (PTS). The privilege token effectively prevents the modifier and owner from colluding to perform unauthorized edits.      
	\item [2.]
	\textbf{Conflict-free multi-level editing.} We subdivide editing privileges into diversified $n$-times Tx-level and Bl-level editing to handle conflicting editing requests and reduce latency, $n$ is defined as the number of edits. Specifically, fine-grained modification privilege tokens are distributed to modifiers with specified credibility levels to perform multi-level editing operations. 
	\item [3.]
	\textbf{Practical Instantiation.} We present the instantiation and security analysis of CDEdit. The implementation and evaluation show that our scheme only needs low performance overhead and can resist conspiracy attacks.	
\end{itemize}.	
	
\section{Overview of CDEdit}
 The CDEdit consists of ciphertext policy attribute-based encryption (CP-ABE) [10], and core mechanisms such as chameleon hashes with ephemeral trapdoor (CHET) and digital signatures(Schnorr signature scheme [15] is used). PTS is responsible for validating and filtering editing requests and issuing appropriate privilege tokens to the modifier group. The CP-ABE encryption algorithm encrypts the ephemeral trapdoor in CHET under the access structure $\mathbb{A}$. If a set of attributes meet the access structure corresponding to the ciphertext, it can be decrypted. Note that  CDEdit is not a centralized control of editing privileges. A valid edited transaction or block must have the following conditions: 1) a token issued by PTS, 2) a valid access attribute policy, and 3) a CH trapdoor key pair.
	
After the owner uploads a variable transaction to the blockchain, the policy-based chameleon hash is output, which contains ciphertext and signature. The upload of immutable transactions is the same as that in traditional blockchain, which we will not discuss more. Considering the different editing needs of different requesters and the convenience of supervising illegal acts, the management and allocation of editing privileges is necessary. In CDEdit, editing privileges controlled and  managed  by classifying it into request privilege, modification privilege and verification privilege. Request privilege requires that legal modifiers have multi-level editing requirements. This scheme requires a certain deposit amount for each editing request to prevent system congestion caused by blocking attacks. Modification privilege is the permission granted by PTS to the modifier group to perform modification. Verification privilege is a license held by all participants in the system. verification privilege provides the transparency and accountability of data on the chain, and effectively identifies abnormal behaviour.
	
Due to the complexity of the actual environment,  we assume that the modifier and the owner entity can intersect. CDEdit provides $n$-times Tx-level edits and $n$-times Bl-level edits for privilege token. Note that  $n$-times edit is defined as allowing $n$-times edits in a reasonable time without multiple token and key generation and verification processes.

\section{Preliminaries}
\subsection{Blockchain Basics}

Using the notation used by [11] and [12], we adopt the following model and notation to describe the blockchain: assume that each block consists of a tuple $B_i= <PreH_i,x_i,ctr_i,r>,i\in [0,N],B_0$ denotes genesis block and $B_N$ denotes the latest block (head of the chain), $PreH_i\in\{0,1\}^K$ denotes the pre-block hash of length $K$ bits, $x_i\in\{0,1\}^*$ denotes the current block data of arbitrary length (include the transaction root hash $TX_{root}$ and others), $ctr_i\in\mathbb{N} $ denotes the randomness $Nonce$ generated by consensus mechanism, and the randomness $r$ for the CH. A block $B_i$ is valid iff
$$
\begin{aligned}
&\text { ValidateBlock }^{D}\left(B_{i}\right): \\
&\qquad=\left(H\left(c t r_{i}, G\left(\operatorname{Pre} H_{i}, x_{i}, r\right)\right)<D\right) \cap\left(c t r_{i}<q\right)=1.
\end{aligned}
$$

Here, $H:\{0,1\}^{*} \rightarrow\{0,1\}^{K}$ and $G:\{0,1\}^{*} \rightarrow\{0,1\}^{K}$ are collision-resistant hash function, they are called the outer hash function and the inner hash function, respectively. The parameter $D\in\mathbb{N}$ is the block's difficulty level, and $q\in\mathbb{N} $ are the maximum number of hash queries in each round of the consensus process.

The rightmost block of blockchain $C$ is called the head of the chain, denoted by $Head(C)$. Any chain $C$ with a head $Head(C):=\left\langle PreH_i,x_i,ctr_i,r\right\rangle $ can be extended to a new longer chain $C':=C||B_{i+1}$ by attaching a valid block $B_{i+1}=\left\langle PreH_{i+1},x_{i+1},ctr_{i+1},r'\right\rangle $ such that $PreH_{i+1}=H(ctr_i,G(PreH_i,x_i,r))$; the head of the new chain $C'$ is $Head(C'):=B_{i+1}$. If $C$ is a prefix of $C'$ we write $C \prec C'$. 

\subsection{Ciphertext-policy attribute-based encryption}
The PCH is constructed from CHET and CP-ABE schemes [13]. The following section formally introduces the concept of access structure associated with ciphertext in CP-ABE, discusses how to encode access structures.

\textbf{Access Structure}. Let $\mathbb{U}$ denote the universe of attributes. A collection $\mathbb{A} \in 2^{\mathbb{U}}\setminus \{\phi\}$ of non-empty subsets is an access structure on $\mathbb{U}$. The subsets in $\mathbb{A}$ are called the authorized sets, and the sets not in $\mathbb{A}$ are called the unauthorized sets. It is called monotone if $\forall B, C \in \mathbb{A}:$ if $B\in \mathbb{A}$ and $B \subseteq C$ the $C \subseteq \mathbb{A}$.

\textbf{Monotone Span Program (MSP)}. As of previous work, linear secret-sharing schemes (LSSS) [14] consist of a share-generating matrix $\mathbb{M}$ with $n_1$ rows and $n_2$ columns which encodes monotone access structures. The monotonic access structure is usually represented as a Boolean formula on attributes with AND and OR operators, which is satisfied if the input attribute computes a value of 1. An alternative way to represent such formulas is to think of access trees. In such a tree, the leaf's form the input attributes, while the inner nodes are associated with the operators AND and OR. LSSS with domain of secrets realizing access structure $\mathbb{A}$ is called linear over $Z_q$ if: 1) The shares of a secret $s\in Z_q$ for each attribute form a vector over $Z_q$, and 2) For $u=\{1,...,n_1\}$, we define a function $\pi$ that maps the row $u$ of $\mathbb{M}$ with attribute $\pi(u)$ from the attribute universe $\mathbb{U}$. Then, the column vector $\vec{v}=\left(s, r_{2}, \ldots, r_{n_{2}}\right)^{T}$, where $s\in Z_q$ is the secret to be shared and $r_{2}, \ldots, r_{n_{2}} \in Z_{q}$ are chosen at random. The $\mathbb{M} \vec{v} \in Z_{q}^{n_{1} \times 1}$ is the vector of $n_1$ shares of the secret $s$ according to LSSS. The share $(\mathbb{M} \vec{v})_{u}$ belongs to attribute $\pi(u)$.

[14] states that every LSSS has the linear reconstruction property. Assume that LSSS is an MSP for the access structure $\mathbb{A}$, $1=\mathbb{A}(\theta)$ is an authorized set and let $I\subset\{1,2,...,n_1\}$ be defined as $I=\left\{u \in\left[n_{1}\right] \cap \pi(u) \in \theta\right\}$. There exist the constants $\left\{\gamma_{\mathrm{u}} \in Z_{q}\right\}_{u \in I}$ such that for any valid share $\left\{\lambda_{u} \in(\mathbb{M} \vec{v})_{u}\right\}_{u \in I}$ of a secret s according to LSSS, $\sum_{u \in I} \gamma_{u} \lambda_{u}=s$. Meanwhile, these constants $\left\{\gamma_{u}\right\}_{u \in I}$ can be found in the polynomial time of the size of the matrix $\mathbb{M}$. For any unauthorized set $\theta^{\prime}$, no such $\left\{\gamma_{u}\right\}$ exist.
\subsection{Policy-Based Chameleon Hashes}
A PCH [13] with message space $\mathcal{M}$ consists of five algorithms $\left\lbrace PPGen,KGen,Hash,Verify,Adapt\right\rbrace $ which are defined as follows.
\begin{itemize}
	\item $\mathcal{PCH}.PPGen\left(1^{\lambda}\right) \rightarrow\left(sk_{PCH}, pk_{PCH}\right)$: On input a security parameter $\lambda\in\mathbb{N}$ in unary, this algorithm outputs the secret key $sk_{PCH}$ and the public key $pk_{PCH}$, where $pk_{PCH}$ is implicitly available to all algorithms.
	\item $\mathcal{P C H}. KGen(sk_{PCH},\theta) \rightarrow sk_\theta$: On input a secret key $sk_{PCH}$ and a set of attributes $\theta \subseteq \mathbb{U}$, the key generation algorithm outputs a secret key $sk_\theta$.
	\item $\mathcal{P C H}. Hash(pk_{PCH},m,\mathbb{A}) \rightarrow(h,r)$:  On input a public key $pk_{PCH}$, access structure $\mathbb{A} \subseteq 2^{\mathbb{U}}$, and a message $m\in \mathcal{M}$, and outputs a hash $h$ and randomness $r$.
	\item $\mathcal{P C H}. Verify(pk_{PCH},m,h,r) \rightarrow b$: On input public key $pk_{PCH}$, a message $m\in\mathcal{M}$, a hash $h$, and a randomness $r$, the algorithm outputs a decision bit $b\in \{0,1\}$.
	\item $\mathcal{P C H}.Adapt(sk_\theta,m,m',h,r) \rightarrow r'$: On input a secret key $sk_\theta$, messages $m,m'\in\mathcal{M}$, hash value $h$ and randomness $r$, and outputs a randomness $r'$.
\end{itemize}
	
Note that we assume that the $KGen$ outputs $\perp$ if $\theta$ is not contained in $\mathbb{U}$ and the $Adapt$ algorithm always verifies if the hash it is given is valid, and output $\perp$ otherwise.

\section{CDEdit System Model}

\subsection{Types of Modification Privilege Token}
	
CDEdit supports four types of tokens with different permission semantics. Each of these different token types corresponds to a collection of four levels of modifiers to enable CDEdit system diversity and anti-collision request control for modification privileges.
\begin{itemize}
	\item One-time Tx-level token ($T_{1tk}$). It represents the simple transaction modification token in the system. A modifier who owns $T_{1tk}$ can only modify a transaction with specific parameters.
	\item $n$-time Tx-level token ($T_{ntk}$). The modifier is allowed to perform $n$-times Tx-level transaction modification operations using the associated parameters or methods before the token expires. Note that $T_{ntk}$ is higher than the permission level of $T_{1tk}$, i.e. the modifier with $T_{ntk}$ can also perform only a single Tx-level modification.
	\item One-time Bl-level token ($B_{1tk}$). It indicates that the modifier performs a modification operation on a whole data block by calling specific parameters. Since each block contains several transactions, $B_{tk}$ has a higher edit permission level than $T_{tk}$.
	\item $n$-times Bl-level token ($B_{ntk}$). It represents the super token in the system, i.e. the highest permission level. A modifier who owns $B_{ntk}$ can replace $n$ consecutive blocks using the associated block parameter.
\end{itemize}
	
We define the sets of modifiers with different credibility levels $\{m_{1T},m_{nT},$ $m_{1B},m_{nB}\}$ respectively, and their modification privileges correspond to the above four types of privileges token, i.e., $\{T_{1tk}\rightarrow m_{1T},T_{ntk}\rightarrow m_{nT},B_{1tk}\rightarrow m_{1B},B_{ntk}\rightarrow m_{nB}\}$. The PTS sets all privileges tokens with an expiration time to avoid repeated invocations by the modifier. Specifically, the modifier can only receive a new token if the editor wants to continue performing edit operations.
	
The modifier sends a privilege token request $req_{tk}=(type||reqPayload)$ to the PTS when it has an edit request, where $reqPayload=(n||ID||index)$, $type\in \{T_{tk},B_{tk}\} $ denotes the type of the edit object, $n$ denotes the number of modifications, $ID$ is the identity of requester, $index$ is the index address of the edit target (transaction or block). PTS receives the request and verifies its validity, and then signs it with the private key $sk_{pts}$, i.e. $Sign_{sk_{pts}}(type || expire||  reqPayload  )$, where $reqPayload$ is an optional field of the token request, whose size varies depending on the $type$. The final return to the modifier is a privilege token
\begin{equation}
pri_{tk}=(type||expire|| index || time||  deposit|| Sign_{sk_{pts}}).
\end{equation}
where $expire$ encodes the expiration time, $deposit$ denotes the information of margin payment, and $time$ is the timestamp of $pri_{tk}$ sent to the modifier.

\begin{figure*}[htbp]
	\centerline{\includegraphics[width=0.7\textwidth]{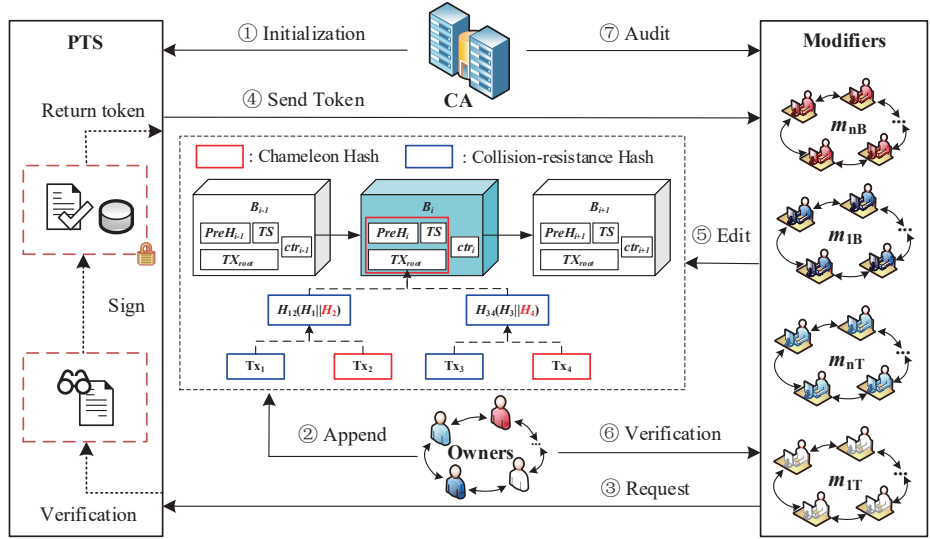}}
	\caption{CDEdit system model.}
	\label{fig1}
\end{figure*}

\subsection{Formal Definitions}\label{AA}
\begin{itemize}
	\item $Setup(1^{\lambda}) \rightarrow(SK,PK)$: It takes a security parameter $\lambda\in \mathbb{N}$ as input, outputs a chameleon hash key pair $(sk,pk)$ and a key pair $(sk_{pts},pk_{pts})$.
	\item $TkGen(sk_{pts},req_{tk},n) \rightarrow pri_{tk}$: It takes a secret key $sk_{pts}$, a editions token request $req_{tk}$ and $n\in \mathbb{N}$, and outputs a privilege token key $pri_{tk}$.
	\item $KeyGen(sk,\theta) \rightarrow sk_\theta$: It takes the secret key set $sk$, where $sk$ includes the secret key $sk_{CHET}$ of CHET and the secret key $msk$ of CP-ABE, and a set of attributes $\theta \in \mathbb{U}$ as input, outputs a secret key $sk_\theta$, which is indexed by an identity of modifier.
	\item $Hash(pk,m,\mathbb{A},ID_j)\rightarrow\left( h,r,\sigma\right) $: It takes the chame-leon pubic key $pk$, a message $m:=\{Tx_{ID},x_i\}\in \mathcal{M}$, an access policy $\mathbb{A}$, and an owner identity $ID_j$ as input, outputs a chameleon hash $h$, a randomness $r$, and signature $\sigma$. Note that $\mathcal{M}=\{0,1\}^*$ denote a general message space, $Tx_{ID}$ and $x_i$ denote the transaction data and $i$-th block data, respectively.
	\item $Verify_m(pri_{tk},pk_{pts}) \rightarrow b$: It takes the privilege token $pri_{tk}$, a public key $pk_{pts}$ as input, and outputs a decision bit $b\in\{0,1\}$.
	\item $Verify(pk,m,h,r,\sigma) \rightarrow b$: It takes the public key $pk$, a message $m$, chameleon hash $h$, randomness $r$, and signature $\sigma$ as input, and outputs a bit $b$.
	\item $Adapt(sk_\theta,m,m',level_{m_i},h,r,\sigma) \rightarrow (r',\sigma')$: It takes the secret key $sk_\theta$, a message $m$, a new message $m'$, index $i$ of the modifier's identity $level_{m_i}$, chameleon hash $h$, randomness $r$, and a signature $\sigma$ as input, outputs a new randomness $r'$ if $1=\mathbb{A}(\theta)$ and a new signature $\sigma'$.
	\item $Audit(sk,m,m',r,r',\sigma,\sigma',h) \rightarrow (level_{m_i}$,$d)$: It takes the chameleon secret key $sk$, messages $m$ and $m'$, randomness $r$ and $r'$, signature $\sigma$ and $\sigma'$, a chameleon hash $h$ as input, and outputs the new privilege level $level_{m_i}$ of the modifier $m_i$, and accountability record $d$.
\end{itemize}
	
Correctness. We say that scheme CDEdit satisfies the correctness property if for all security parameters $\lambda\in \mathbb{N}$, for all $(sk,pk)\leftarrow Setup\left(1^{\lambda}\right)$, for all $\theta \in \mathbb{U}$, for all $pri_{tk} \leftarrow TkGen(sk_{pts},pk_{pts},req_{tk},n)$, for all $\theta \in \mathbb{A}$, for all $sk_\theta \leftarrow KeyGen(sk,\theta)$, for all $m\in \mathcal{M}$, for all $(h,r,\sigma) \leftarrow Hash(pk,msk,m,\mathbb{A},ID_j)$, for all $m'\in \mathcal{M}$, we have for all $(r',\sigma') \leftarrow Adapt(sk_\theta,m,m',level_{m_i},h,r,\sigma)$, that $1=Verify_m(pri_{tk},pk_{pts})$ and $1=Verify(pk,m,h,r,\sigma)=Verify(pk,m',h,r',\sigma')$.
	
\subsection{System Model}
The CDEdit system model contains four main types of entities: central authority (CA), PTS, modifiers, and owners, as shown in Fig.\ref{fig1}. Assume that the transactions $Tx_2 $ and $Tx_4$ in boxed red are policy-based mutable transactions with different access permissions, and that they can be modified without changing the hash value. The transactions $Tx_1$ and $Tx_3$ framed in blue are immutable transactions generated by the owner using a collision-resistant hash. This $TX_{root}$ is the root hash of the Merkle tree, which accumulates all the transactions within the block. Thus, the $i$-th blocks is a mutable block $B_i=\left\langle PreH_i,x_i,ctr_i,r\right\rangle $ generated by PCH, where $PreH_{i+1}=H(ctr_i,PCH(PreH_i,x_i,r))$. These system components are described below.

\begin{itemize}
	\item CA represents the administrator in the permissioned blockchain and is responsible for managing the PTS and system initialization (see \ding{172}). CA broadcasts public parameters at system initialization and audits the modifier's editing behaviour after receiving a report message from the owner (see \ding{178}).
	\item Modifier has the permission to send edit requests to the PTS (see \ding{174}), and the permission to modify the transactions or block data (see \ding{176}). A group of modifiers is classified into different modifier sets $level_{m_i}$ by credibility level, i.e., a single Tx-level modifier set $m_{1T}$, a multiple Tx-level modifier set $m_{nT}$, a single Bl-level modifier set $m_{1B}$ and a multiple Bl-level modifiers set $m_{nB}$ (credibility level from small to large $m_{1T}<m_{nT}<m_{1B}<m_{nB}$).
	\begin{figure}[htbp]
	    \centerline{\includegraphics[width=0.4\textwidth]{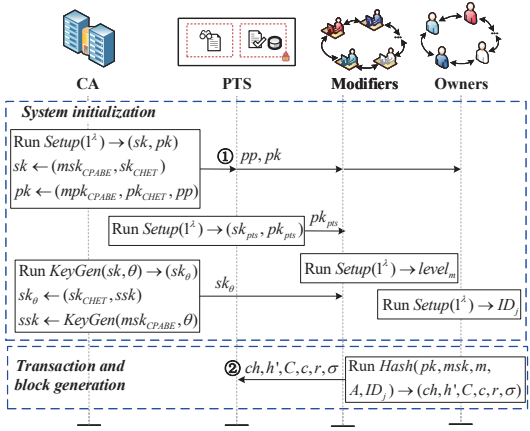}}
	    \caption{System initialization, transaction and block generation.}
        \label{fig2}
    \end{figure}
    \begin{figure}[htbp]
	    \centerline{\includegraphics[width=0.4\textwidth]{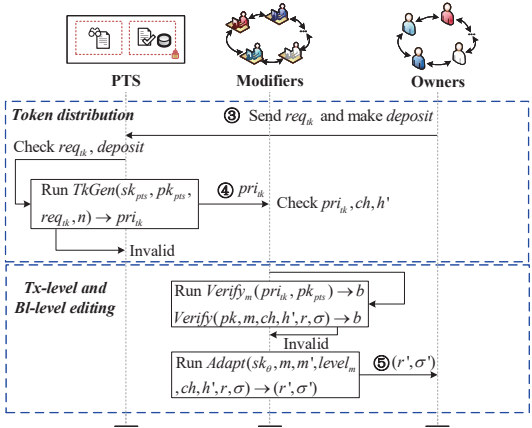}}
	    \caption{Token distribution, and Tx-level and Bl-level editing in CDEdit.}
	    \label{fig3}
    \end{figure}
	\item PTS is responsible for verification of edit requests from modifiers and providing the corresponding edit privilege token (see \ding{175}) to the modifier, i.e., assigning modification privileges. A privilege token determines the editing scope of a particular modifier set. Note that the sender of an edit request does not necessarily belong to the same level of modifier group. 
	\item The owner represents the primary participant and is responsible for adding immutable or mutable transactions (see \ding{173}). The owner can also send edit requests based on modification requirements, as well as validate the editing behaviour of specific modifiers (see \ding{174}).
\end{itemize}

\subsubsection{System Initialization.} This phase can be divided into system setup, PTS setup, modifier setup and owner setup, as shown at the top of Fig.~\ref{fig2}.
	
\begin{itemize}
	\item \emph{System setup}. CA chosen a parameter $\lambda$, and outputs key pair $\left( sk,pk\right) $ by running the algorithm $Setup\left( 1^\lambda\right)$ (see \ding{172}), where $sk=\left( msk_{CP-ABE},sk_{CHET}\right)$, $pk=( mpk_{CP-ABE}$, $pk_{CHET}, pp)$, and $pp\leftarrow Setup_{\mathcal{DS}}(1^\lambda)$, $( msk, mpk) $ is generated by $Setup_{CP-ABE}\left( 1^\lambda\right) $, while $(sk_{CHET},$ $ pk_{CHET})$ is generated by $Setup_{CHET}(1^\lambda)$. CA selects the CH secret key $sk$ and the attribute set $\theta$, and outputs the secret key $sk_\theta \leftarrow (sk_{CHET},ssk)$ by $KeyGen(sk,\theta)$ algorithm, where $ssk\leftarrow KeyGen(msk_{CP-ABE},\theta)$. Finally, the CA broadcasts the public parameter $pp$ and the chameleon public key $pk$ to the network.
	\item \emph{PTS setup}. A trusted PTS is specified by CA and generates its own key pair $(sk_{pts},pk_{pts})$ by running the $Setup(1^\lambda)$ algorithm. $(sk_{pts},pk_{pts})$ is used for subsequent PTS issuance and signing of the token.
	\item \emph{Modifier setup}. After receiving $pp, pk$ from CA, the transaction modifier runs $Setup(1^\lambda)$ algorithm to generate its own identity information (or level) $level_m$. The initial level is set by CA, such as $m_{1T}$ or $m_{1B}$.
	\item \emph{Owner setup}. After receiving $pp, pk$ from the CA, the transaction owner runs the $Setup(1^\lambda)$ algorithm to obtain its own identity address information $ID_j$.
\end{itemize}
	
\subsubsection{Transaction and Block Generation.} In CDEdit, mutable transactions and mutable blocks are allowed to be generated, as shown at the bottom of Fig.~\ref{fig2}.
\begin{itemize}
	\item \emph{Generate mutable transactions}. owner executes the $Hash(pk,msk,m,p,ID_j)$ to generate a PCH with message $m$ and upload it to the chain (see \ding{173}). Concretely, a hash $ch$, a randomness $r$ and ephemeral trapdoor $etd$ are generated by the $Hash(pk,m)$ algorithm in CHET, and bring $etd$ into the $Enc(mpk,R,\mathbb{A},ID_j)$ algorithm of CP-ABE to generate the ciphertext $C$. Then, the signed message $c$ generated using $\mathcal{D S}.KeyGen(pp,sk,etd)$ algorithm is signed to obtain $\sigma$.
	\item \emph{Generate mutable block}. The $TX_{root}$ is generated by several transactions through multiple hashing algorithms, and the timestamp $TS$ constitutes the block data $x_i$. Generate a mutable block $B_i=(PreH_i,x_i,ctr_i)$ by taking $x_i$ and the pre-block hash $PreH_i$, and the randomness $ctr_i$.
\end{itemize}
	
\subsubsection{Token Distribution.} This phase is where the modifier obtains the token issued by the PTS, as shown at the top of Fig.~\ref{fig3}. First, the modifier initiates an edit request $req_{tk}$ to the PTS and pays a certain amount of deposit (see \ding{174}). After receiving $req_{tk}$, the PTS checks whether the deposit paid satisfies the requested edit type cost, as well as verifies the corresponding parameter information and signature $\lambda_{pts}$. Then, PTS generates the corresponding privilege token $pri_{tk}$ by $TkGen(sk_{pts},pk_{pts},req_{tk},n)$ algorithm, where $n$ indicates the number of modifications. Finally, PTS distributes $pri_{tk}$ (see \ding{175}) to the set of modifiers of the relevant level $level_{m_i}$.
	
\subsubsection{Tx-level and Bl-level editing.} As shown at the bottom of Fig.~\ref{fig3}, the modifier group $level_{m_i}$ receives the token $pri_{tk}$ and needs to check its validity and execute the $Verify_m ((pri_{tk},pk_{pts})$ algorithm to verify the token type and the signature $\sigma_{pts}$ of the token. Only $pri_{tk}$ is valid to continue the next operation. Otherwise, stop. Validating the token before the actual edit execution can effectively filter out invalid requests. Then choose to perform the following edit types (see \ding{176}).
	
\begin{itemize}
	\item \emph{Tx-level edit}. Let $m$ as transaction $Tx_{ID}$, the modifier $level_m$ satisfying the set of attributes $S\in \mathbb{A}$ starts by executing the $Verify(pk,Tx_{ID},ch,h',r,\sigma)$ algorithm to verify the validity of the hash pair $(ch,h')$ and runs the adaption algorithm $Adapt(sk_\theta,Tx_{ID},Tx_{ID}',level_m,ch,h',r,\sigma)$ to output the new randomness $r'$ and the signature $\sigma'$.
	\item \emph{Bl-level edit}. Let $m$ as data $x_i$ in $i$-th mutable block, finding a hash collision by the $Adapt(sk_\theta,x_i,x_i',level_m,ch,h',r,\sigma)$ algorithm, and output $(r',\sigma')$. Bl-level editing is similar to Tx-level editing, where block $B_i$ changes to $B_i'$.
\end{itemize}	
\begin{figure}[htbp]
	\centerline{\includegraphics[width=0.5\textwidth]{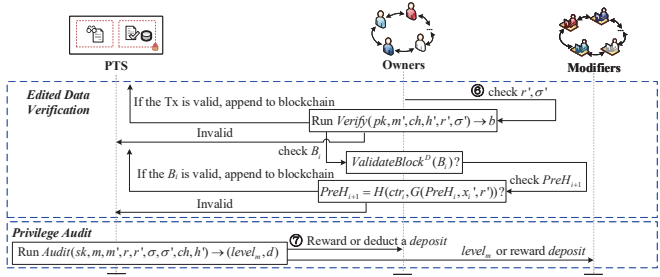}}
	\caption{Edited data verification and Privilege audit in CDEdit.}
    \label{fig4}
\end{figure}
\subsubsection{Edited Data Verification.} As shown at the top of Fig.~\ref{fig4}, there are two types of Tx-level and Bl-level editing in the CDEdit. Therefore, we also consider two kinds of data validation processes: mutable transaction verification and mutable block verification.
	
\begin{itemize}
	\item \emph{Mutable transaction verification}. Each participant verifies the validity of the received mutable transaction by executing the $Verify$ $(pk,Tx',ch,h',r',\sigma')$ algorithm with the chameleon public key $pk$, the message $m'$ and the CH key pair $(ch,h')$. To reduce the verification overhead, the transaction owner is mainly responsible for the verification operation in CDEdit. If the transaction is valid, the owner will modify the local copy of the transaction, otherwise rejected.
	\item \emph{Mutable block verification}. First, the owner runs the $Verify(pk,x_i',ch,h',r',\sigma')$ algorithm to verify the validity of the data $x_i$ in the mutable block. Second, validate the validity of a mutable block by $ValidateBlock^D(B_i)$, where $D$ is the block's difficulty level. Third, verify whether the equation $PreH_{i+1}=H(ctr_i,PCH(PreH_i,x_i',r'))$ holds by $PreH_i$ in the current mutable block.
\end{itemize}
\subsubsection{Privilege Audit.} Assume that the owner verifies that the modifier has malicious behaviour, such as performing more than $n$-times edits or stealing higher-level tokens, and the owner sends a report message to the CA, as shown at the bottom of Fig.~\ref{fig4}. The CA receives a report from the owner, it executes the $Audit(sk,m,m',r,r',\sigma,\sigma',ch,h')$ algorithm to audit the modifier for malicious operations. If it has illegal behaviour, CA will reduce the $level_{m_i}$ of the modifier or even kick out the modifier group and reward part of the $deposit$ to the owner. Otherwise, CA will increase the credibility level of the modifier.

\subsection{Secure Model}

In the CDEdit system, we assume that the CA and PTS are fully trusted and that owners and modifiers are untrusted participants. Owners and modifiers may launch a conspiracy attack where they try to come to perform unauthorized editing operations. Furthermore, CDEdit considers four security properties, including indistinguishability, collision resistance, and EUF-CMA security.

\begin{itemize}
	\item \emph{Indistinguishability}. Informally, indistinguishability requires that the adversary cannot distinguish whether the randomness $r$ of the $Hash$ is new or was created using an $Adapt$ algorithm. 
	\emph{Definition} 4.1. A CDEdit scheme is indistinguishable if for any PPT adversary $\mathcal{A}$ in the IND experiment $Exp_{A,\mathcal{CDE}dit}^{IND}(\lambda)$, $Adv_{A,\mathcal{CDE}dit}^{IND}(\lambda)$ is negligible in $\lambda$.	
	$$
	\operatorname{Adv}_{\mathcal{A},\mathcal{CDE} d i t }^{\mathrm{IND}}(\lambda):=\mid \operatorname{Pr}\left[\operatorname{Exp}_{\mathcal{A}, \mathcal{CDE} d i t }^{\mathrm{IND}}(\lambda)=1\right]-1 / 2 \mid.
	$$
	\item \emph{Collision Resistance}. Informally, collision resistance indicates that an insider with a secret key associated with some of the attributes cannot discover the collision of hashes associated with the policy. Also, they do not find the collision of hashes related to the key mismatching policy by $KeyGen$ oracle. 
	\emph{Definition} 4.2. A CDEdit scheme is collision resistance if for any PPT adversary $\mathcal{A}$ in the CR experiment $Exp_{A,\mathcal{CDE}dit}^{CR} (\lambda)$, $Adv_{A,\mathcal{CDE}dit}^{CR} (\lambda)$ is negligible in $\lambda$.
	$$
	\operatorname{Adv}_{\mathcal{A}, \mathcal{CDE} d i t }^{\mathrm{CR}}(\lambda):= \operatorname{Pr}\left[\operatorname{Exp}_{\mathcal{A}, \mathcal{CDE} d i t }^{\mathrm{CR}}(\lambda)=1\right].
	$$
	\item \emph{EUF-CMA Security}. The signature scheme $\mathcal{D S}$ is called Existential Unforgeability Against Adaptive Chosen Messages Attacks, referred to as EUF-CMA security.
	\emph{Definition} 4.3. A CDEdit scheme is EUF-CMA secure if for any PPT adversary $\mathcal{A}$, the following advantage is negligible in $\lambda$.
	$$
	\operatorname{Adv}_{\mathcal{A}, \mathcal{CDE} d i t }^{\mathrm{EUF-CMA}}(\lambda):= \operatorname{Pr}\left[\operatorname{Exp}_{\mathcal{A}, \mathcal{CDE} d i t }^{\mathrm{EUF-CMA}}(\lambda)=1\right].
	$$
\end{itemize}

\section{Instantiation and Security Analysis}
\subsection{Instantiation}

To construct an efficient and secure CDEdit scheme, we rely on the latest CP-ABE scheme [10], known as FAME, and the CHET scheme [16]. The FAME uses a hash function $H_1$, which maps arbitrary binary strings to elements of the group $\mathbb{G}$. In CDEdit, two types of inputs are given to a hash function $H_1$: inputs of the form $(x,\ell,t)$ or that of the form $(v,\ell,t)$, where $x$ is an arbitrary string, $v$ is a positive integer, $\ell \in \{1,2,3\}$ and $t\in \{1,2\}$. Then, we represent these two inputs as $x\ell t$ and $0v\ell t$, respectively. Where 0 is used to distinguish between these two strings. Moreover, we define the identity of all participants as vectors and assume that the identity of the owner $ID_j=(I_1,\cdots,I_j)\in (\mathbb{Z}_q )^j$, the transaction modifier identity is $level_{m_i}=(I_1,\cdots,I_i)\in (\mathbb{Z}_q )^i$, where $j\leq i$. 
\begin{itemize}
	\item $Setup(1^\lambda)$: It inputs a security parameter $\lambda$ as input and obtain $(q,\mathbb{G},\mathbb{H},\mathbb{G}_T,\widehat{e},g,h)$, where $g$ is the generator of group $\mathbb{G}$, $h$ is the generator of group $\mathbb{H}$, $(\mathbb{G}, \mathbb{H}, \mathbb{G}_T)$ is groups of order $q$. Pick $(a_1,a_2,b_1,b_2,x,y)\leftarrow \mathbb{Z}_q^*$, $\left\lbrace z_1,\cdots,z_k\right\rbrace \leftarrow \mathbb{Z}_q$, $(d_1,d_2,d_3)\leftarrow \mathbb{Z}_q$, and $d=$ $d_1+d_2+d_3$, to calculate  $H_1=h^{a_1}$,$H_2=h^{a_2}$, $T_1=\widehat{e}(g,h)^{d_1\cdot a_1+d_3}$, $T_2=\widehat{e}(g,h)^{d_2\cdot a_2+d_3}$.Then, it outputs a master public key $mpk=(g,h,H_1,$ $H_2,T_1,T_2,\{g^{z_1},\cdots,g^{z_k}\},\{h^{z_1 },\cdots,h^{z_k}\})$ and a master secret key $msk=(a_1,a_2,b_1,b_2,g^{d_1},g^{d_2},g^{d_3},\{z_1,\cdots,$ $z_k \})$, chameleon key pair $(sk,pk)=(x,h^x)$, and PTS key pair $(sk_{pts},pk_{pts} )=(y,\widehat{e}(g,h)^y)$. We define a secret credential as $ID_j=\prod_{i=0}^jh\cdot h_{k-i-1}^{I_i}$ for each user.
	
	\item $TkGen(sk_{pts},pk_{pts},req_{tk},n)$: It inputs an edit request $req_{tk}=(type||n||ID_j||index)$, $n\in \mathbb{N}$, $type$ and $index$ are arbitrary strings, to calculate $\sigma_{pts}=k+sk_{pts}\cdot \mathcal{H}(pk_{pts}||req_{tk}||kg||deposit))$, where $kg=k\cdot g$, $k\leftarrow \mathbb{Z}_q^*$, and outputs a privilege token $pri_{tk}=(req_{tk},\sigma_{pts},kg,time)$, where $time$ is the given period.
	
	\item $KeyGen(sk,\theta)$: It inputs a chameleon secret key $sk$, and a set of attributes $\theta $, and picks $R\leftarrow \mathbb{Z}_q$, $(r_1,r_2)\leftarrow \mathbb{Z}_q^*$ and $r=r_1+r_2$, computes $sk_0=(h^{b_1\cdot r_1 },h^{b_2\cdot r_2},h^r,g^R)$. For all $y\in \theta$ and $t=\{1,2\}$, picks $\sigma_y\leftarrow \mathbb{Z}_q$, compute
	$sk_{y,t}=H_1(y1t)^{(b_1\cdot r_1)/(a_t)}\cdot H_1(y2t)^{(b_2\cdot r_2)/(a_t)}\cdot H_1(y3t)^{(r_1+r_2)/(a_t)}\cdot g^{(\sigma_y)/(a_t)}$, and set $sk_y=(sk_{y,1},sk_{y,2},g^{-\sigma_y})$. Then, it picks $\sigma'\leftarrow \mathbb{Z}_q$, for $t=\{1,2\}$, computes
	\begin{equation}
	sk_t'=g^{d_t}\cdot H_1(011t)^{\frac{b_1\cdot r_1}{a_t}}\cdot H_1(012t)^{\frac{b_2\cdot r_2}{a_t}} \cdot H_1(013t)^{\frac{r_1+r_2}{a_t}}\cdot g^{\frac{\sigma'}{a_t}}.
	\end{equation}
	and sets $sk'=(sk_1',sk_2',g^{d_3},g^{-\sigma'})$. Then, it computes $sk_1=g^d\cdot level_{m_i}^r\cdot g^R$, $sk_2=\{g_{i-1}^r,\cdots,g_1^r\}$. Last, outputs a secret key $sk_{\theta_i}=(x,ssk_i)$, where $ssk_i=(sk_0,\{sk_y\}_{y\in \theta},sk',sk_1,sk_2)$, a modifier's identity in the decryption key as $level_{m_i}=\prod_{j=0}^ig\cdot g_{k-j-1}^{I_j}$.
	
	\item $Hash(pk,m,\mathbb{M},ID_j)$: To hash a message $m$ under a policy $(\mathbb{M},\pi)$, and an identity $ID_j$, an owner performs the following. 1) compute $p=pk^r$, where $r\leftarrow \mathbb{Z}_q^*$ is a randomness; 2) compute ephemeral trapdoor $etd=H_2(R)$, $h'=h^{etd}$. Note that R denote a short bit-string, $R\leftarrow \mathbb{Z}_q^*$; 3) compute a CH $ch=p\cdot h'^m$; 4) generate a verification key pair as $(sk,vk)=(s, {ID_j}^s)$, where $s=s_1+s_2$,$(s_1,s_2)\leftarrow \mathbb{Z}_q^*$; 5) generate a ciphertext on the message $M=(r,R)$ with the policy $(\mathbb{M},\pi)$ and identity $ID_j$, compute $ct_0=(H_1^{s_1}$,$H_2^{s_2}$,$h^s)$. Then, for $i=\{1,\cdots,n_1\}$ and $\ell=\{1,2,3\}$, compute
	\begin{equation}
	\begin{aligned}
	ct_{i,\ell}&=H_1(\pi(i)\ell 1)^{s_1}\cdot H_1(\pi(i)\ell 2)^{s_2}\cdot {\prod_{j=1}^{n_2}}[H_1 \\
	&\quad (0j\ell 1)^{s_1}\cdot H_1(0j\ell2)^{s_2}]^{(\mathbb{M})_{i,j}}. 
	\end{aligned}
	\end{equation}
	where, $(\mathbb{M})_{i,j}$ denotes the $(i,j)$-th element of $\mathbb{M}$. Then, it computes $ct=r\oplus G(T_1^{s_1}\cdot T_2^{s_2})$, $ct'=R\oplus H_2(\widehat{e}(g,h^d)^s)$, $ct_1=ct_2={ID_j}^s$, $ct_3=ct_1^s$, and $C=(ct_0,\{ct_i\}_{i\in n_1},ct,ct',ct_1,ct_2,ct_3)$. 6) compute signed message $c=h^{sk+etd}$, and generate a signature $\sigma=esk+sk\cdot H_2(epk||c)$, where $epk=g^{esk}$ and $(epk,esk)$ denotes an ephemeral key pair. Eventually, it output $(m,p,h',ch,C,c,epk,\sigma)$.
	
	\item $Verify_m (pri_{tk},pk_{pts})$: PTS has to verify the validity of the $pri_{tk}$ before proceeding with the modification operation. It inputs a $pri_{tk}$ and $pk_{pts}$, and outputs 1 if $\sigma_{pts}\cdot g=kg+pk_{pts}\cdot H_1(pk_{pts}||req_{tk}||kg||deposit)$.
	
\begin{figure*}[htbp]
	\centering
	\subfigure[]{
		\begin{minipage}[t]{0.3\linewidth}
			\centering
			\includegraphics[width=0.9\textwidth]{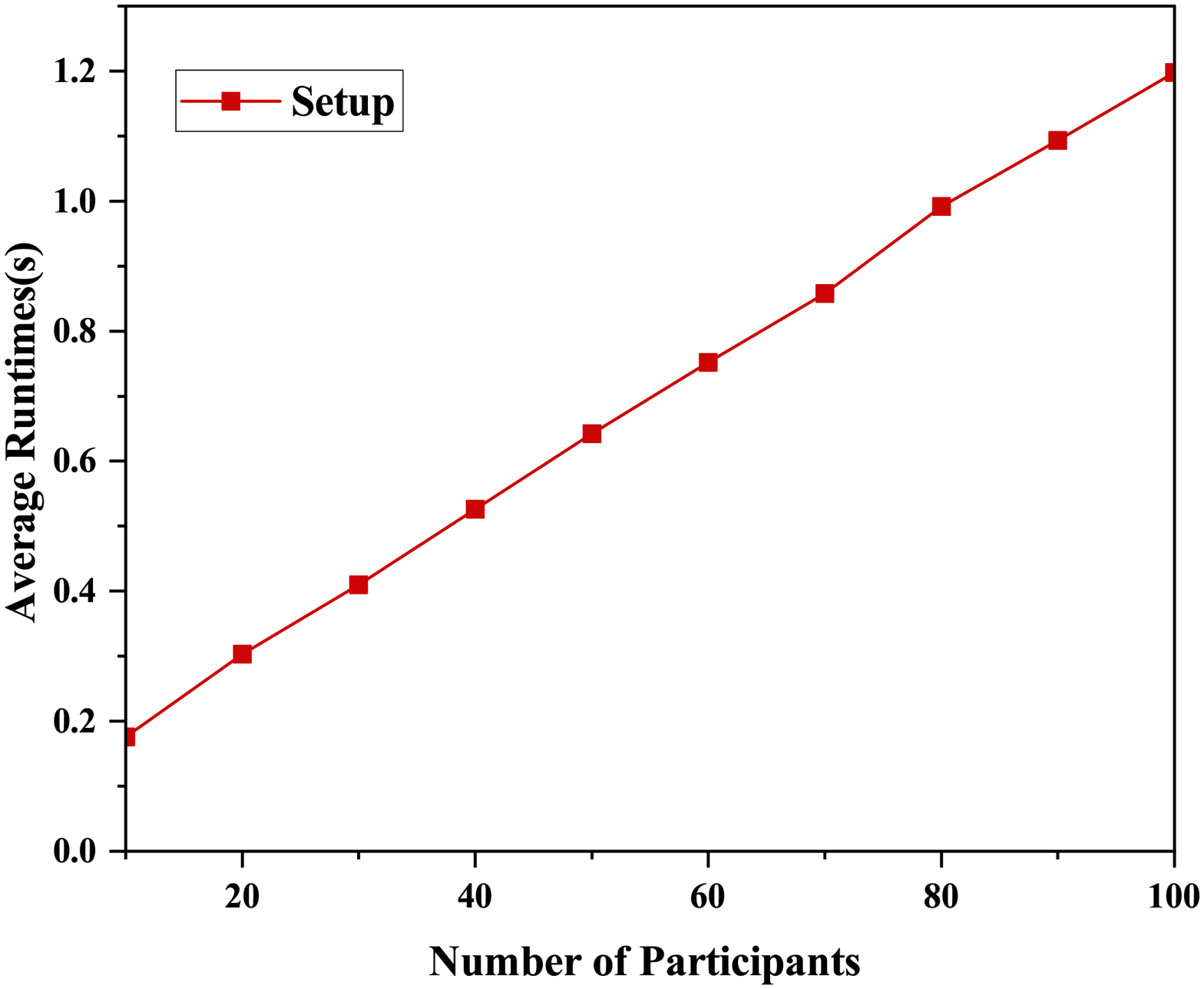}
			\label{fig5-a}
		\end{minipage}%
	}%
	\subfigure[]{
		\begin{minipage}[t]{0.3\linewidth}
			\centering
			\includegraphics[width=0.9\textwidth]{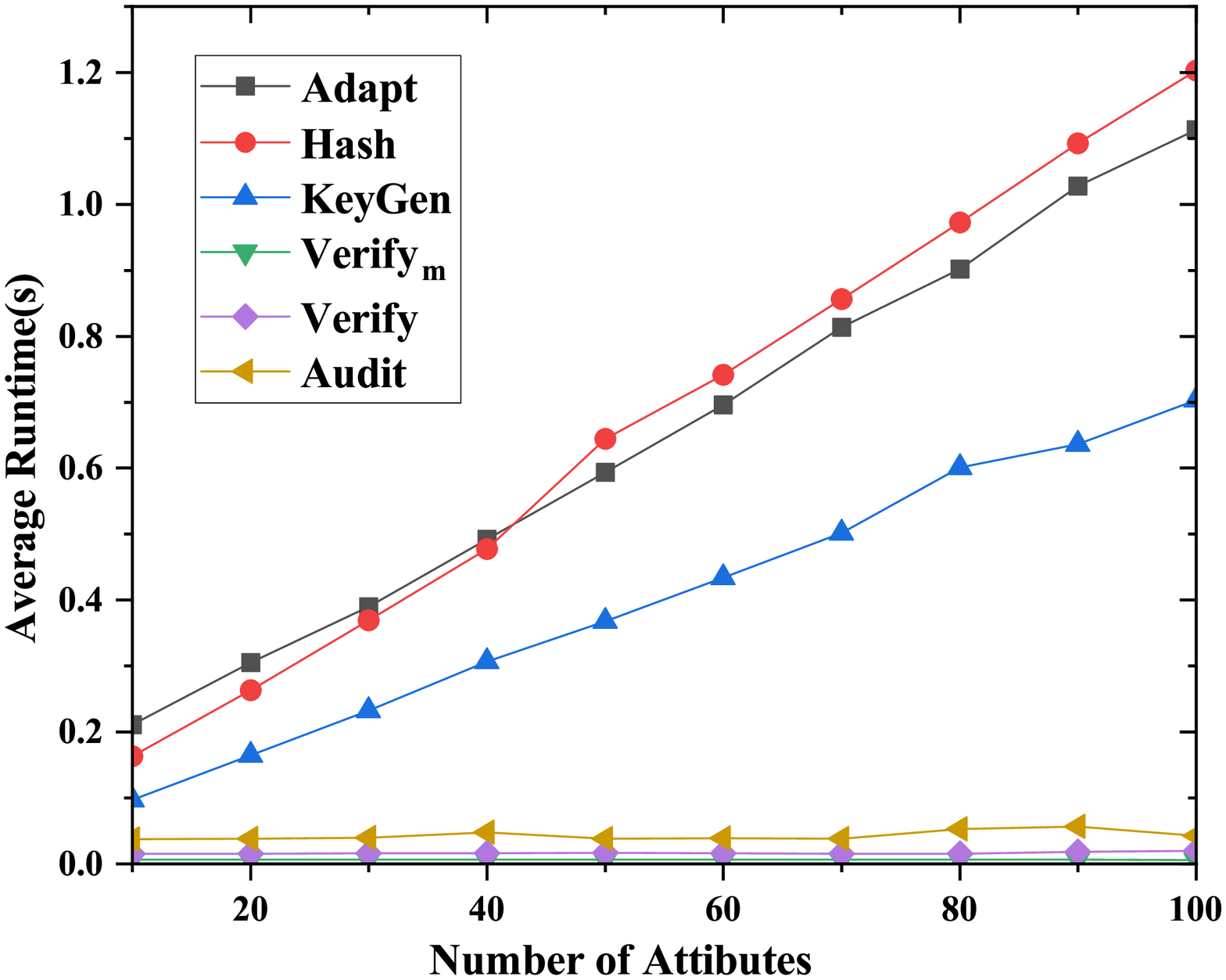}
			\label{fig5-b}
		\end{minipage}%
	}%
	\subfigure[]{
		\begin{minipage}[t]{0.3\linewidth}
			\centering
			\includegraphics[width=0.9\textwidth]{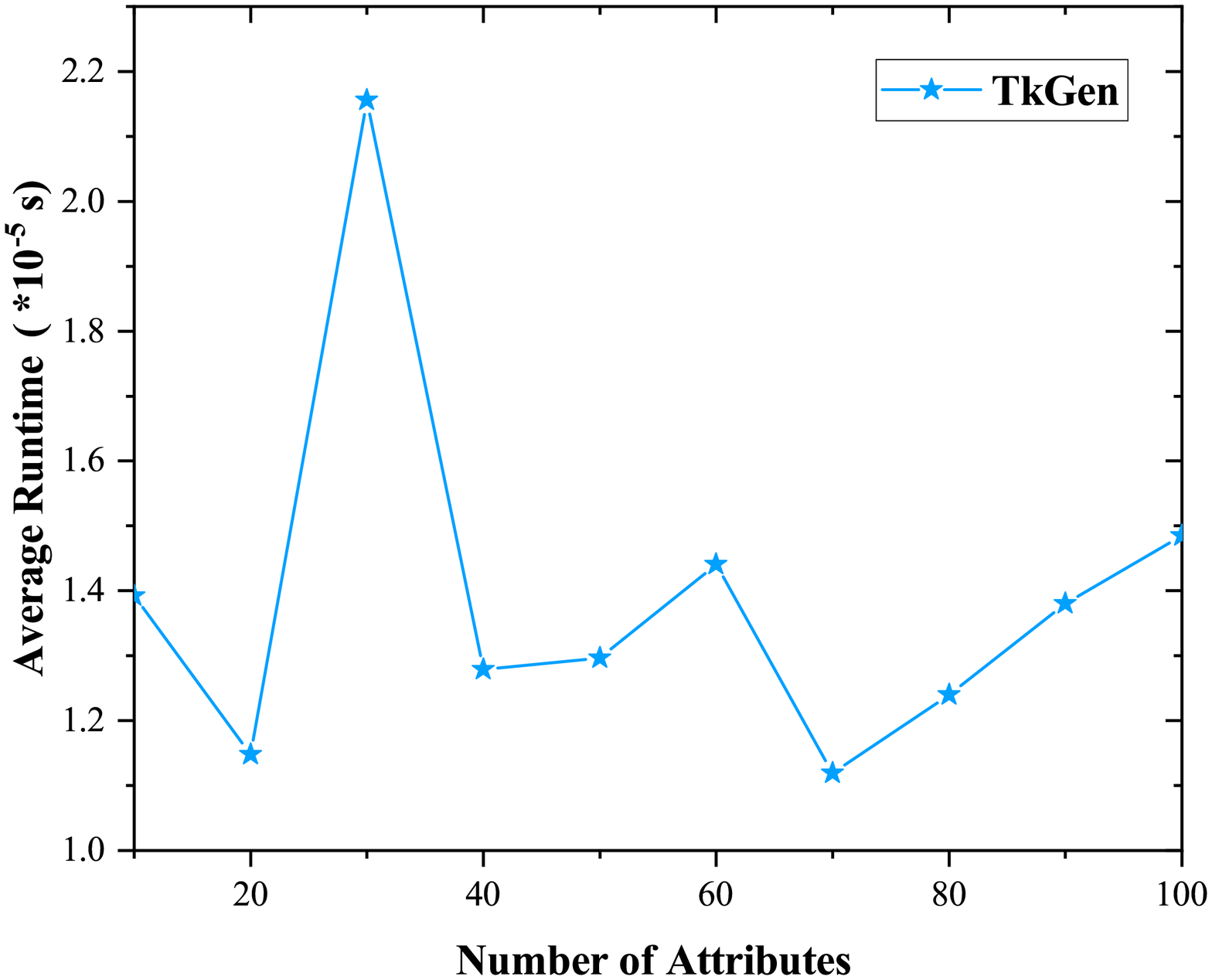}
			\label{fig5-c}
		\end{minipage}
	}%
	\centering
	\caption{The average runtime of system's algorithms.}
\end{figure*}
	
	\item $Verify(pk,m,p,h',ch,C,c,epk,\sigma)$: Each user can verify whether a given hash $(ch,h')$ is valid, it outputs 1 if $ch=p\cdot h'^m$ and $\widehat{e}(g,ct_2)^\sigma=\widehat{e}(epk,ct_1)\cdot \widehat{e}(g,ct_3 )^{H_2(epk||c)}$, where $ct_1=ct_2={ID_j}^s$.
	\item $Adapt(sk_{\theta_i},m,m',p,h',ch,C,c,epk,\sigma,level_{m_i})$: The modifier who is granted edit privileges performs the following operations with the secret key $sk_{\theta_i}$, a new message $m'$ (if selected as Bl-level edit then $m'=TX_{root}||TS$), and an identity $level_{m_i}$; 1) verify that $Verify(pk,m,p,h',ch,C,c,epk,\sigma)$ is equal to 1; 2) For obtain the encrypted randomness $r$, and assuming that attributes $\theta$ in $ssk_i$ satisfies the MSP $(\mathbb{M},\pi)$, then compute coefficients $\{\lambda_i\}_{i\in I}$. Now, compute
	\begin{equation}
	\begin{aligned}
	num &=\widehat{e}(\prod_{i\in I}ct_{i,1}^{\lambda_i},sk_{0,1})\cdot \widehat{e}(\prod_{i\in I}ct_{i,2}^{\lambda_i},sk_{0,2})\\
	&\quad \cdot \widehat{e}(\prod_{i\in I}ct_{i,3}^{\lambda_i},sk_{0,3}).
	\end{aligned}
	\end{equation}
	\begin{equation}
	\begin{aligned}
	den &=\widehat{e}(sk_1'\cdot \prod_{i\in I}ct_{\pi(i),1}^{\lambda_i},ct_{0,1})\cdot 
	\widehat{e}(sk_2'\cdot \prod_{i\in I}ct_{\pi(i),2}^{\lambda_i},ct_{0,2})\\ 
	&\qquad \cdot \widehat{e}(sk_3'\cdot \prod_{i\in I}ct_{\pi(i),3}^{\lambda_i},ct_{0,3}).
	\end{aligned}
	\end{equation}
	and output $r=ct\oplus G(den/num)=ct\oplus G(T_1^{s_1}\cdot T_2^{s_2})$. Here $sk_{0,1}$, $sk_{0,2}$, $sk_{0,3}$ denote the first, second and third elements of $sk_0$, and the same for $ct_0$. 3) derive a new randomness $r'=r+(m-m')·etd/x$, and compute $p'=pk^{r'}$, where $etd=H_2(R)$. 4) generate a verification key pair as $(sk',vk'):=(s',{ID_j}^{s'})$, where $s'=s_1'+s_2'$,$(s_1',s_2')\leftarrow \mathbb{Z}_q^*$. 5) generate a ciphertext $C'$ on message $M'=(r',R)$ using randomness $(s_1',s_2')$, under policy $(\mathbb{M},\pi)$ and identity $level_{m_i}$. 6) compute signed message $c'=h^{sk'+etd}$, and generate a signature $\sigma'=esk'+sk'\cdot H_2 (epk'||c')$, where $epk'=g^esk'$. Eventually, it output $(m',p',h',ch,C',c',epk',\sigma')$.
\end{itemize}

\emph{Correctness}. We use the $Audit$ algorithm to ensure the accountability of CDEdit system. All users can publicly verify the correctness of a new version of a transaction or block. The final audit is performed by CA and includes the following: 1) inputs $(m,m',p,p',h',ch,C',c',epk',\sigma,\sigma')$, and verify CH $ch=p\cdot h'^{m}=p'\cdot h'^{m'}$. 2) verify message signature pair $(c,\sigma)$ under $(epk,vk)$, and $(c',\sigma')$ under $(epk',vk')$. 3) verify the number of edits. Eventually, it output new $level_{m_i}$ and accountability record $d$. 

\subsection{Security Analysis}
The owners and modifiers cannot edit without a CA-issued privilege token, even if they decrypt an ephemeral trapdoor, which means this scheme resists a conspiracy attack. As detailed in Appendix, we have performed a security analysis of the proposed scheme.

\emph{Theorem} 5.1. If the CDEdit scheme is based on an indistinguishable CHET, then the CDEdit scheme is indistinguishable.
	
\emph{Theorem} 5.2. If the CDEdit scheme is based on collision resistant CHET and an IND-CCA2 secure ABE, then the CDEdit scheme is collision resistant.
	
\emph{Theorem} 5.3. If the digital signature is EUF-CMA secure, then the proposed CDEdit scheme is EUF-CMA secure.

\section{Implementation and Evaluation}
To demonstrate the performance of our proposed CDEdit system, we implement it in Python 3.6.9 using Charm 0.43 framework [17] and on a PC running 64-bit Ubuntu 18.04 LTS with Intel Core i5(1.60GHz$\times$4) and 7.8GiB RAM. We use MNT224 curve [18] for pairing because it is the best Type-III curve in PBC, and it has around 96-bit security level [19]. Note that the CDEdit scheme does not change the chain length but replaces the original blocks with new blocks. Therefore, the mining difficulty and consensus algorithm (e.g., PBFT) of the actual blockchain is not used as direct influencing factors. We simulated the latency time of the CDEdit system for the different number of participants and sets of attributes. The implementation code is available on GitHub [20].
\renewcommand\arraystretch{1.25}
\begin{table}[htbp]
	\caption{AVERAGE RUNNING TIME OF TX-LEVEL AND BL-LEVEL EDITS}
	\begin{threeparttable}
	\begin{tabular}{|c|c|c|}
		\hline
		Type & Edit Times & Average Runtime \\
		\hline 
		\multirow{2}{*}{Tx-level} & $n$ one-time &\parbox[c]{4cm}{$t_{set}+n(t_{tk}+t_{key}+t_{h}+t_{ver}+t_{ver_m}+t_{ad}+t_{au})$} \\
		\cline{2-3}
		& $n$-times &\parbox[c]{4cm}{$t+(n-1)(t_h+2·t_{ver}+t_{ad}+t_{au})$}  \\
		\hline 
		\multirow{2}{*}{Bl-level} 	& $n$ one-time &\parbox[c]{4cm}{$t_{set}+n(t_{tk}+t_{key}+t_{ver_m}+t_{ad}+t_{au})$} \\
		\cline{2-3}
		& $n$-times &\parbox[c]{4cm}{$t+(n-1)(t_{ver}+t_{ad}+t_{au})$}  \\
		\hline	
	\end{tabular}
	\label{tab1}
	\footnotesize
	$^{*}$ $t_{set}$, $t_{tk}$, $t_{key}$, $t_h$, $t_{ver}$, $t_{ver_m}$, $t_{ad}$, $t_{au}$ denote the runtime of algorithms $Setup$, $TkGen$, $KeyGen$, $Hash$, $Verify$, $Verify_m$, $Adapt$, $Audit$, respectively. $t$ denotes the total runtime of all algorithms.
	\end{threeparttable}
\end{table}
\begin{figure}[htbp]
	\centerline{\includegraphics[width=0.4\textwidth]{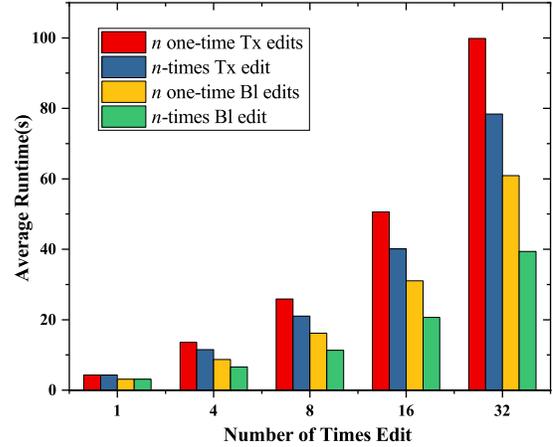}}
	\caption{Runtime comparison for diversified editing.}
	\label{fig6}
\end{figure}

We applied CDEdit to an identity management scenario of IoT devices in a region. We tested the performance of a set of participating devices and attributes of $10, 20, \cdots, 100$, respectively. It is shown experimentally that the runtime of $Setup$ algorithms increases linearly with the increase in the number of devices and demonstrated in Fig.~\ref{fig5-a}. The average running time is only 1.2s even with 100 devices, which is acceptable for practical applications. Further, we fixed the number of devices to 120 and then experimented with $KeyGen$, $Hash$, and $Adapt$ algorithms to evaluate the relationship between attribute size and average running time. From Fig.~\ref{fig5-b}, it can be seen that the performance of these algorithms shows a linear relationship with the number of attributes or the policy size. Even if the number of attributes is 100, the running time of $KeyGen$ algorithm is only 0.7s, $Adapt$ algorithm is only 1.11s, and $Hash$ algorithm is only 1.2s. We also evaluated each validation algorithm to demonstrate performance benefits. As the number of attributes increases, the running time of $Verify_m$ algorithm is maintained between (0.0136, 0.042)ms, $Verify$ algorithm is preserved between (0.155, 0.198)ms, and $Audit$ algorithm is supported between (0.34, 0.36)ms. We have experimented with $TkGen$ algorithms that run in an average time of about (1.1, 2.2) $\times10^{-5}$s and shown in Fig.~\ref{fig5-c}. As a result, the running time of each algorithm of the CDEdit system is acceptable with its availability. We conducted contrasted experiments with $n$ one-time edits and $n$-times edits to implement multi-level editing operations for privilege tokens in our scheme, respectively. We set the attribute size to 100, and collated the average running time of Tx-level and Bl-level edits in $n$ one-time edits and $n$-times edits as shown in Table I.

Finally, we set $n$ to 1, 4, 8, 16, 32, respectively. We discuss the influence of the number of edits on the running time of Tx-level and Bl-level edits, as shown in Fig.~\ref{fig6}. In Bl-level editing, it is assumed that a complete chain is already available, regardless of the runtime of block generation. The $n$-times edit has less runtime than $n$ one-time edits because there is no need to compute additional setup, token and keygen algorithms per round. And the advantage becomes more and more obvious as the number of edits increases.

\section{Related Work}
Since the concept of redactable blockchain was proposed [12], exploring the management and distribution of editing privileges has always been the focus of researchers. Ashritha [21] et al. proposed a secret-sharing based redactable blockchain scheme that implements the editing operation by securely distributing CH trapdoor keys to a predefined group of verifiers and through a voting consensus process. Nevertheless, complex cryptographic tools to manage CH trapdoor keys are inefficient when the group of verifiers is large. Dealer et al. [13] introduced a PCH function to control modifications in a permissioned blockchain by CHET [16] at the fine granularity of trapdoor keys to the editor's privilege. Any participant in CHET who has a long-term trapdoor and an ephemeral trapdoor can quickly compute the new randomness $r'$. But, PCH is exposed to the risk of conspiracy attacks, where the modifier and owner can collide with each other to find a set of attributes that satisfy.

\textbf{Concurrent and Independent Work}. Puddu et al. [22] have proposed $\mu$-chain with mutable transactions, whose editing privileges are determined by the policy constructed from the sender. However, transactions in the $\mu$-chain are encrypted, which reduces the audibility of the data written. Deuber et al. [23] proposed a redactable blockchain with consensus-based voting in a permissionless blockchain. The edit operation is executed whenever and only when the user's edit request collects enough votes from the miner. Similarly, Thyagarajan et al. [24] proposed a compatible Reparo generic protocol and instantiated it on typical blockchain applications like Ethereum and Bitcoin. Li et al. [25] proposed a redactable blockchain with an instant editing function in the permissionless blockchain, where the blockchain editing is decided jointly by selecting a committee vote. Unfortunately, several of the above redactable blockchains consensus-based voting suffer from defects such as too long voting cycles, poor scalability, and inefficiency. If the edit request cost is low, the system is vulnerable to DDoS attacks and Sybil attacks [26]. We list the comparison of utility features in the current major solutions shown in Table~\ref{tab2}.
\renewcommand\arraystretch{1.25}
\begin{table*}[htbp]
	\caption{COMPARISON AMONG CURRENT REDACTABLE BLOCKCHAIN SOLUTIONS AND OURS}
	\begin{center}
		\begin{tabular}{|c|c|c|c|c|c|c|}
			\hline
			\textbf{Schemes}&\textbf{Type} &\textbf{Core Mechanism}&\textbf{Redaction}&\textbf{$n$-time}&\textbf{Verification}&\textbf{Collusion Resistant} \\
			\hline 
			Ateniese et.al. [12] & Permissioned & CH,PKI & Bl-level & $\times$ & $\times$ & $\times$ \\
			\hline
			Deuber et.al. [23]& Permissionless & Consensus/Voting & Tx-level & $\times$ & $\checkmark$ & N/A  \\
			\hline 
			PCHBA [6] & Permissioned & PCH, Signatues & Tx-level & $\times$ & $\checkmark$ & $\checkmark$ \\
			\hline
			KERB [7] & Permissioned & CH, Signatues & Tx-level & $\checkmark$ & $\checkmark$ & $\checkmark$  \\
			\hline
			CDEdit & Permissioned & PCH, Signatues,PTS & Multi-level & $\checkmark$ & $\checkmark$ & $\checkmark$  \\
			\hline
			\multicolumn{4}{l}{$^{*}$Multi-level: Supporting both Tx-level and Bl-level. N/A: Not applicable.}	
		\end{tabular}
		\label{tab2}
	\end{center}
\end{table*}

\section{Conclusion}
In this work, we proposed a highly applicable redactable blockchain with controllable editing privilege and diversified editing types. The proposed scheme supports multi-level editing at both Tx-level and Bl-level in blockchain and effectively prevents malicious editing and conspiracy attacks. The instantiation shows that the framework can be extended to other permissioned blockchain applications. In future work, we aim to further improve system performance and reduce performance overhead, as well as extend to permissionless blockchain systems.

\section*{Appendix}
\emph{Proof of Theorem} 5.1. The reduction is executed between an adversary $\mathcal{A}$ and a simulator $\mathcal{S}$. Assume that adversary $\mathcal{A}$ can guess the CH value with a nonnegligible probability in the proposed CDEdit. Let simulator $\mathcal{S}$ denote a distinguish against CHET, who is given a chameleon public key $pk^*$ and a HashOrAdapt oracle, aims to break the indistinguishability of CHET. Let $q$ be an upper bound on the queries made to the $Hash$ oracle. $\mathcal{S}$ randomly chooses $i\in \{1,\cdots,q\}$ as a guess for the index of the HashOrAdapt query. Then, the distinguisher $\mathcal{S}$'s challenger directly hashes a message $Hash(pk^*,m)\rightarrow (h,r)$.

\begin{itemize}
	\item Setup: $\mathcal{S}$ generates $n$ owners and corresponding identity $\{ID_j\}$, and sets up the game for $\mathcal{A}$.
	\item Challenges: $\mathcal{S}$ randomly selects an owner as of the attribute policy and sets its chameleon public key to $pk^*$ . $\mathcal{S}$ can honestly generate a decryption key for any transaction modifier associated with that $ID_j$ and attribute set $\theta$. If $\mathcal{A}$ submits a tuple $(m_o,m_1,\mathbb{A})$ in the $i$-th query, then $\mathcal{S}$ randomly choose $b\in \{0,1\}$, and obtains the CH $(h_b,r_b)$ from the HashOrAdapt oracle on the message $(m_0,m_1)$. Further, according to the protocol, $\mathcal{S}$ simulates the message signature pair $(c,\sigma)$ and the ciphertext $C$. 
	\item Guess: Finally, $\mathcal{S}$ returns $(h_b,r_b,\sigma_b)$ to $\mathcal{A}$. $\mathcal{S}$ follows the results of $\mathcal{A}$ output. If $\mathcal{A}$ guesses the random bit correctly, $\mathcal{S}$ can break the indistinguishability of CHET.\\
\end{itemize}

\emph{Proof of Theorem} 5.2. We prove the theorem in a sequence of Game $i\in\left\lbrace 0,1,2,3\right\rbrace$, denoting by $Pr[S_i]$ the success probability of the adversary $\mathcal{A}$ in Game $i$. Assume that the queries number to the $Hash'$ oracle be denoted by at most $q$.

\begin{itemize}
	\item Game 0: This is the original $CR$ security experiment for collision resistant.
	\item Game 1: As Game 0, but $\mathcal{S}$ randomly guess the index $i^*$ corresponding to the $Hash'$ oracle which return the CH $(h^*,C^*)$ which will be attacked by the adversary. We store the hash $(h^*,C^*)$ as well as the corresponding randomness $r^*$ and the ephemeral trapdoor $etd^*$. If during the simulation we detect that the guess is wrong, we will abort. Otherwise the same winning probability in Game 1 as in Game 0, and has $Pr[S_1]=Pr[S_0]/q$.
	\item Game 2: As Game 1, but whenever $\mathcal{S}$ receives an adapt query for a hash $(h,C)$, where $C=C^*$ and not decrypt, but directly adapt using $etd^*$. The same winning probability in Game 2 as in Game 1 under the perfect correctness of the encryption scheme, and has $Pr[S_2]=Pr[S_1]$.
	\item 	Game 3: As Game 2, but $\mathcal{S}$ changes the simulation of the $Hash$ algorithm within the $i^*$-th query to the $Hash_{CHET}'$ oracle, and run $C\leftarrow \prod_{ABE}Enc(0^{|etd|},\mathcal{A})$ and locally store $etd$. 
\end{itemize}

We claim that Game 2 and Game 3 are indistinguishable under the IND-CCA2 security of $\prod_{ABE}$, i.e., $|Pr[S_3]-Pr[S_2]|\neq Adv_{,ABE}^{IND-CCA2}(\lambda)$. Next, we show that we can use an adaptive IND-CCA challenger to effectively interpolate between Game 2 and Game 3. Get $mpk$ from the IND-CCA challenger, make $msk\rightarrow \bot$, and continue honestly with other setups. Then, the respective oracles provided by the challenger are used to simulate queries to the key generation oracle. After the $i^*$-th query to $Hash'$ oracle, output $(etd,0^{|etd|} ,\mathcal{A},state)$ to the challenger to obtain $(C^*,state)$, and set $C\rightarrow C^*$. Further, for adaptive queries with hash returned at the $i^*$-th query to $Hash'$, we directly use $etd$ for adapt without prior decryption. Note that once it turns out that our guess of the index $i^*$ is wrong, this ensures that we never have to answer queries which involve queries against the challenger's oracle. This means that we can simulate Game 2 if the challenger chooses $b=0$, and we can simulate Game 3 if $b=1$. \\

\emph{Proof of Theorem} 5.3. Let simulator $\mathcal{F}$ be given a public key $pk^*$ and a signature oracle $\mathcal{O}^{Sign}$ used to forge DS with the aim of breaking the EUF-CMA security of $\mathcal{DS}$. $\mathcal{F}$ randomly chooses a CH, and setup its verification key as $pk^*$. Then, $\mathcal{F}$ randomly chooses $i\in \{1,2,\cdots,q\}$ as a guess for the index of the forgery with respect to that CH. $\mathcal{F}$ can honestly completes the remainder of $Setup$. $\mathcal{F}$ obtains a signature $\sigma$ from his signature oracle $\mathcal{O}^{Sign}$. $\mathcal{F}$ honestly generates the CH and the ciphertext and returns $(h,r,\sigma,C,c)$ to adversary $\mathcal{A}$. Due to the homomorphism of $\mathcal{DS}$, the message-signature pair as well as the verification key can be perfectly modeled by $\mathcal{F}$ for any adaptive query, and additionally the ephemeral trapdoor etd is chosen by $\mathcal{F}$. $\mathcal{F}$ records all the simulated CH in a set Q. When forging attack occurs, i.e., $\mathcal{A}$ outputs $(h^*,r^*,\sigma^*,C^*,c^*)$, $\mathcal{F}$ check the following conditions:
\begin{itemize}
	\item The forgery attack occurs on the $q$-th guess;
	\item The ciphertext $C^*$ encrypts the ephemeral trapdoor $etd$;
	\item The message-signature pair $(c^*,\sigma^*)$ links to $pk^*$.
	\item The message-signature pair $(c^*,\sigma^*)\notin Q$.
	\item $1\leftarrow Verify(vk^*,c^*,\sigma^*)$ and $1\leftarrow Verify$ $(h^*,m^*,r^*)$. 	
\end{itemize}

\end{document}